# Radiative decay effects influence the local electromagnetic response of the monolayer graphene with surface corrugations in terahertz range

Yu.A. Firsov[1], N.E. Firsova[2,3]

[1]A.F.Ioffe Physical-Technical Institute, the Russian Academy of Sciences, St.Petersburg, 194021, Russia,
**email:** yuafirsov@rambler.ru
[2]Institute for Problems of Mechanical Engineering, the Russian Academy of Sciences, St. Petersburg 199178, Bolsoy Prospect V.O. 61, Russia,
[3]State Polytechnical University, St. Petersburg 195231, Politechnicheskaya 29, Russia,
**email:** nef2@mail.ru

**Abstract**

We continue the study of surface corrugations influence on the monolayer graphene local electromagnetic response in terahertz range we started earlier. The effects of radiative decay, double-valley structure of charge carriers spectrum in graphene and the breathing surface curvature form induced synthetic electric fields are taken into account. To fulfill this program the generalized nonlinear self-consistent equation is obtained. In case of weak external alternating electric field $\vec{E}^{ext}(t) = \vec{E}_0 \cos \omega t$ for the obtained equation in linear approximation on the external electric field the exact solution is found. It shows that in this case we get local dephasing in induced current paths depending on the surface form $z = h(x,y)$. This theoretical result qualitatively explains the corresponding experiments with local current patterns in graphene without using fully quantum approach which is necessary for theoretical description of such phenomenon in graphene nanoribbons. Besides the formulae obtained in the present paper could become the basis of the new method of the imaging of surface corrugations form for given picture of local current paths. The obtained results allow to study mechanical behavior of materials at nanoscale deviates from macroscoping established concepts and this is of particular importance for graphene.

## 1. Introduction

Single carbon atom sheet was made in 2004 in [1] and first 2D crystal in 2005 [2]. The first non trivial experimental observations ("Dirac-like" charge carriers spectrum and anomalous quantum Hall effect) were published in 2005 [2-3]. Within almost 10 years much more new exciting facts concerning graphene properties were elucidated. One of the most important of them is that strictly speaking graphene membrane is not "flatland" (2D system) as it was declared at the beginning.

Its surface is deformed by out–of-plane corrugations ("ripples") as it was observed for the first time in [4,5]. The in-plane deformations mutual interaction influence save membrane from being crumpled (see for instance [6]). It was shown 70 years ago by Peierls [7-8] and Landau [9] that strictly speaking 2D crystal can not exist since thermal fluctuations should destroy long range order because of thermal fluctuations. Out-of-plane displacements i.e. the surface form $z = h(x,y)$ and in-plane strains create the gauge field which were studied in [10], [11]. The influence of these static pseudomagnetic fields the physical properties of graphene was studied in details: their connection with minimal conductivity [12], weak antilocalization [13], topological insulator state [14], pseudomagnetic quantum Hall effect [14] and so on. All these results as well and other ones of the sort obtained up to 2007 are discussed in interesting review [15].

If out-of-plane corrugations are excited by an external a.c. electric field they begin to move ("breathe") and an additional ("extra") synthetic electric field $\vec{E}_{syn}(x,y,t) = -\frac{1}{c}\frac{\partial \vec{A}}{\partial t}(x,y,t)$ will arise (here $\vec{A}(x,y,t)$ is vector potential corresponding to "internal" gauge field (see above)). As far as we know the influence of $\vec{E}_{syn}(x,y,t)$ in graphene was considered in [16] and by us [17]. In [16] it was shown that (f-ph) damping rate was determined by interaction with these fields due to surface deformation by moving flexural phonons and its time derivative. In [17] we have shown that synthetic fields, excited by time-dependent electromotive force in a graphene nanoresonator, and currents generated by them lead to the new loss mechanism which essentially influences the quality factor value. In [18] we studied the total induced current (from both valleys) in irradiated corrugated graphene membrane taking into account synthetic electric fields generated by activating external periodic electric fields with due regard of the inversion and time-reversal symmetry breaking [19]. The "breathing" surface form $z = h(x,y,t)$ determines the synthetic electric field dependence $\vec{E}_{syn}(x,y,t)$ on the point $(x,y)$ in every moment. The formula for $\vec{E}_{syn}(x,y,t)$ was obtained under assumption that the external electric field is weak enough to consider the spectrum to be continuous (Stark levels are washed out due to current carriers scattering and thermal fluctuation) and suggesting the "breathing" membrane linear dependence on the external electric field. Under these assumptions the current paths depending on the point were found for given surface form and its curvature. So the synthetic field $\vec{E}_{syn}(x,y,t)$ differs by strength, direction and phase and consequently current paths are different and their directions and phases change in neighborhood of every ripple.

Actually it means that we have another model of disordered system where the curvature also depends on time i.e. disorder is not stable (see below "Discussion and Conclusion"). In the present paper we are going to study in addition the role of radiative decay of charge carriers "overheated" by terahertz irradiation which may be more effective than standard relaxation mechanism and make the picture of special alternating current paths more exact.

In [18] we did not take into account e-e interaction considering graphene response to homogeneous radiation in terahertz range. There are many cases (see for instance the review [20] and discussion below), when in the 2D system of interacting Dirac electrons many-body effects should be taken into account. However sometimes this interaction does not play significant role (see below). For instance it was shown that together with quasiparticles ("massless electrons" and "holes") an important role may play plasmons (collective charge density excitations) and "plasmarons" [21] ("dressed" quasiparticle exitations coupled to plasmons). These predictions are based on a very specific form of quasiparticle spectral function obtained theoretically [22-23] and observed experimentally by variety of direct and indirect methods: electron energy-loss spectroscopy (EELS),(see in [24]), angle-resolved photoemission spectroscopy (ARPES) [21], scanning tunneling spectroscopy (STS)

[24] on exfoliated graphene sheets and on epitaxial graphene samples.

Large efforts were made to observe these predictions in optical measurements. As a rule, indirect methods are used "by engineering plasmons coupling to infrared light in a number of intriguing ways", [20], by making more strong light-matter interaction and alleviating the momentum mismatch between plasmon modes and of the incident radiation.

Thus we come to conclusion that under incident homogenous radiation surface plasmons (SP) cannot be driven without using special efforts: either by using AFM tip to confine the incident radiation to $q$ nanoscale region around the tip as Z.Fei, D.N.Basov et al [26] or by exciting SP in graphene micro-ribbon array or by using stacked graphene microdiscs to realize electromagnetic radiation shield with 97,5% effectiveness [20].

The manifestation of e-e interaction in optical and other properties in graphene considered in more details in our previous paper [18]

To conclude the discussion about the strong or weak influence of e-e interactions in optical experiments in monolayer graphene we also refer to three papers [27-29] (here [27] is a theoretical article and [28-29] are experimental ones sent in 2014). In [27] it is written that the negative dynamical conductivity discovered before in experiment one can explain only taking into account e-e interactions. In [28] it is suggested with reasonable arguments that the mentioned phenomenon should be explained by defect mediated collision of the hot carriers with the acoustic phonons. The mentioned experimental data once again confirm our opinion described above that without special serious tricks e-e interactions may not manifest (display) itself.

In [29] a transient decrease in graphene conductivity was observed. The THz frequency-dependence of graphene photoconductive response increases in the Drude weight.

It should be noted that recently the study of graphene current patterns under the action of the electric field became a vogue. The case of the constant electric field and the flatland model was theoretically analyzed on the basis of quantum description and non-equilibrium Keldysh-Green formalism in [30] and using the trajectory-based semiclassical analysis in [31]

In [18] we were the first to consider the case with time-dependent electric field and the monolayer graphene membrane having corrugations. We also took into account the double valley spectrum. We suggested that the external electric field is weak enough to consider the spectrum as continuous (see above) and the formula for synthetic electric field was obtained suggesting the "breathing" linear dependence on the external electric field. Under these assumptions the local current paths depending on the point were found for given surface form and its curvature.

Summarizing all above mentioned it may be told that we carried on our studies of out-of-plane corrugations (ripples) influence on monolayer graphene electromagnetic response based on the kinetic Boltzman equation for Fermi-Dirac momentum distribution function and linear Dirac energy spectrum ($\sim k$) for current carriers, but taking into account double-valleys picture of energy spectrum in first Brillouin zone for Dirac electrons. We took into account inversion and time-reversal symmetry broken by ripples in graphene membrane. The same way as in our paper [18] we did not take into account "direct" e-e interactions (a detailed explanation is given in [18]) but important terms for instance radiation rate which is proportionel to $e^2$ (here $e$ is the electron charge) arise due to the using of the very specific "self-consistent-field approach developed in [32] and [33] which takes into account mutual influence of excited and induced currents.

Note that experimental research of local charge transport visualized with atomic resolution using conductive atomic force microscopy [34] showed the current patterns dependence on the point. We think that this dependence is connected with corrugations which are always present and this experiment proves the correctness of our theory. There are also other experimental results of the sort (see [35]…) which used a new technique that allows one to probe transport by creating a movable scatter with an SPM tip.

Our theoretical description is based on the kinetic equation and quantum description of Dirac charge carriers with double valleys spectrum.

In the present paper we get the nonlinear self-consistent equation with corrugations taken into account and we take into consideration radiation loss.

The time-dependent electric current $j(x, y, t)$ excited in the sample under the action of external a.c. electric field $\vec{E}_{ext}(t)$ creates in its turn a secondary (induced) electric field $\vec{E}_{ind}(t)$ which acts back on electrons and should be added to the external field. Our results obtained below for corrugated (rippled) graphene membrane are compared with the results obtained in [32], [33] for flatland model of the graphene membrane.

We also consider the case when the external field is weak enough so that this nonlinear equation turns to be linear. In this case we obtain its exact solution which shows that in this case the currents get dephasing which depends on the point. This dephasing is determined by the surface curvature in the point. So observing the dephasing map we can get information about the surface form of the membrane. It gives the idea of the new method of imaging of ripples which could be useful for many applications.

The adequate methods for theoretical description of current paths structure for the given surface form and experimental efforts to find a more precise (exact) methods for current paths imaging is concerned to be one of the most actual problems to be solved for graphene. It would be also of utmost importance for the further development of graphene based nanoelectronics and DNA sequencing [36].

A more detailed discussion of all these items will be given in conclusion.

## 2. Generalized self-consistent equation

To describe the electromagnetic response or total current in corrugated graphene membrane in linear approximation taking into account the radiative loss we should find the self-consistent equation generalizing the one obtained in [33] for flatland model. In [18] we obtained the kinetic equation in the form

$$\frac{\partial f(\vec{p}, \vec{r}, t)}{\partial t} - e[\vec{E}^{ext}(t) + \vec{E}^{syn}(\vec{r}, t)]\frac{\partial f(\vec{p}, \vec{r}, t)}{\partial \vec{p}} = 0 \qquad (1)$$

where synthetic electric field $\vec{E}^{syn}(\vec{r}, t)$ appears as a consequence of the presence of out-of plane deformations $z = h(x, y)$. Here

$$\vec{E}^{syn} = -c^{-1} \partial \vec{A}(x,y,t)/\partial t \qquad (2)$$

and $\vec{A}(x,y,t)$ is a gauge field. We used for it the formula obtained in [11]

$$A_x = -\frac{1}{2} A^0 \left[(h_{xx})^2 - (h_{yy})^2\right] a^2, \quad A_y = A^0 \left[h_{xy}(h_{xx} + h_{yy})\right] a^2 \qquad (3)$$

$$A^0 = 3/4 \cdot \epsilon_1/e \cdot c/v_F \qquad (4)$$

Here $\epsilon_1 = 2{,}89 ev$, $\vec{K} = a^{-1}(4\pi/3\sqrt{3}, 0)$ corresponds to a Dirac point

Assuming that the external electric field is weak enough to use the linear approximation $\Delta a(t) = \eta(x,y) E_0 \sin \omega t$ (we consider for simplicity that $\eta(x,y) = const$) it was obtained in [18]. To simplify the form of the equations we obtain we shall assume below that $\eta(x,y) = const$

$$(\vec{E}^{syn})_x (x,y,t) = -E^0(\omega) E_{0x} I_x(x,y) \sin \omega t, \qquad (5)$$

$$(\vec{E}^{syn})_y (x,y,t) = -E^0(\omega) E_{0y} I_y (x,y) \sin \omega t \qquad (6)$$

where

$$E^0(\omega) = 3/4 \cdot \epsilon_1/e \cdot \omega/V_F, \quad I_x = a\eta(h_{xx}^2 - h_{yy}^2), \quad I_y = 2a\eta(h_{xx} + h_{yy})h_{xy}, \qquad (7)$$

The Eq(1) for the temperature close to zero has the exact solution

$$f(\vec{p}, \vec{r}, t) = F_0(|\vec{p} - \vec{p}_0(\vec{r}, t)|), \qquad F_0(\vec{p}) = \left[1 + exp\left(\frac{vp - \mu}{kT}\right)\right]^{-1} \qquad (8)$$

where $F_0(p)$ is the Fermi-Dirac distribution function, and

$$\vec{p}_0(\vec{r}, t) = -e \int_{-\infty}^{t} \left[\vec{E}^{ext}(t') + \vec{E}^{syn}(\vec{r}, t')\right] dt' \qquad (9)$$

If the temperature is zero, $T = 0$, and the chemical potential is finite, $\mu > 0$, the current $\vec{j}(\vec{r}, t)$ can be presented in the form

$$j_x = \frac{g_s e v_F}{(2\pi\hbar)^2} p_F^2 G_x(Q_x, Q_y) \frac{2 P_{0x}}{\sqrt{1 + P_{0x}^2 + P_{0y}^2}}, j_y = \frac{g_s e v_F}{(2\pi\hbar)^2} p_F^2 G_y(Q_x, Q_y) \frac{2 P_{0y}}{\sqrt{1 + P_{0x}^2 + P_{0y}^2}} \qquad (10)$$

where

$$G_x(Q_x, Q_y) = \frac{1}{Q_x} \int_0^{\pi/2} \left[\sqrt{1 + Q_y \sin \varphi + Q_x \cos \varphi} - \sqrt{1 + Q_y \sin \varphi - Q_x \cos \varphi}\right] \cos \varphi d\varphi$$

$$G_y(Q_x, Q_y) = \frac{1}{Q_y} \int_0^{\pi/2} \left[\sqrt{1 + Q_x \cos \varphi + Q_y \sin \varphi} - \sqrt{1 + Q_x \cos \varphi - Q_y \sin \varphi}\right] \sin \varphi \, d\varphi$$

$$Q_{x,y} = \frac{2P_{0x,0y}}{1 + P_{0x}^2 + P_{0y}^2}, \qquad P_{ox,oy} = \frac{p_{0x,0y}}{p_F} \qquad (11)$$

The equation (1) and its solution (8), (9) is the generalization of kinetic equation considered in [33] for flatland. Note that according to the Landau-Peierls theorem demonstrated in [7-9] these corrugations exist always and consequently our generalization is essential.

All the formulae written above were obtained without taking into account the radiative loss. Now we should find the effect produced by this circumstance. We assume that the external electromagnetic field is incident upon graphene membrane along z-axis and induces the ac current in the layer. In general however the timedependent electric current creates in its turn a secondary (induced) electric field $\vec{E}^{ind}$ which acts back on the electrons and should be added to the external field. Calculating response of the system one should take into account that electrons respond not to the external but to the total self-consistent electric field

$$\vec{E}^{tot} = \vec{E}^{tot}_{ext}(t) + \vec{E}^{tot}_{syn}(\vec{r},t), \qquad \vec{E}^{tot}_{ext}(t) = \vec{E}_{ext} + \vec{E}^{ind}_{ext}, \qquad \vec{E}^{tot}_{syn}(t) = \vec{E}_{syn} + \vec{E}^{ind}_{syn}$$

So instead of the equation (1) for the momentum distribution function of the electron we obtain

$$\frac{\partial f(\vec{p},\vec{r},t)}{\partial t} - e\big[\vec{E}^{tot}_{ext}(t) + \vec{E}^{tot}_{syn}(\vec{r},t)\big]\frac{\partial f(\vec{p},\vec{r},t)}{\partial \vec{p}} = 0 \qquad (12)$$

The solution of this equation can be written again in the form

$$\vec{p}_0 = -e \int_{-\infty}^{t} \vec{E}^{tot}_{z=0}(t')dt' \qquad (13)$$

where the field $\vec{E}^{tot}_{z=0}$ is not known and should be calculated self-consistently. To do this we recall that the current and the electric field are related by the Maxwell equations [37]

$$\vec{E}^{ind}_{z=0} = -\frac{2\pi \vec{j}}{c} \qquad (14)$$

Contributing (13), (14) (12) we get the following self-consistent equation of motion for the momentum $\vec{p}_0$

$$\frac{d}{dt}\vec{p}_0 + e\frac{2\pi}{c}[\vec{j}^{ext} + \vec{j}^{syn}] = e[\vec{E}^{ext}(t) + \vec{E}^{syn}(\vec{r},t)] \qquad (15)$$

Introducing $\vec{P} = -\vec{p}/p_F$ and $\tau = \omega t$ we can rewrite (15) in the form

$$\frac{d}{d\tau}P_{0x} + \frac{\gamma}{\omega}\frac{P_{0x}}{\sqrt{1 + P_{0x}^2 + P_{0y}^2}}G_x(Q_x, Q_y) = \frac{e}{\omega p_F}\big[E_x^{ext}(t) + E_x^{syn}(\vec{r},t)\big] \qquad (16)$$

$$\frac{d}{d\tau}P_{0y} + \frac{\gamma}{\omega}\frac{P_{0y}}{\sqrt{1 + P_{0x}^2 + P_{0y}^2}}G_y(Q_x, Q_y) = \frac{e}{\omega p_F}\big[E_y^{ext}(t) + E_y^{syn}(\vec{r},t)\big] \qquad (17)$$

$$\gamma = \frac{g_s}{\pi}\frac{e^2}{\hbar c}\frac{v_F p_F}{\hbar} = v_F \frac{e^2}{\hbar c}\sqrt{g_s \pi n_s} \qquad (18)$$

Here $\gamma$ has physical meaning of the radiative decay rate. For experimentally relevant densities $n_s \approx 10^{12}$ we have $\gamma \approx 7THz$. After this nonlinear equation should be resolved with respect to the momentum $\vec{p}_0$ and the current can be found from the Eqs (10), (11)

Eqs (16)-(18) describe the nonlinear self-consistent response of graphene to an arbitrary external time-dependent electric field $\vec{E}^{ext}_{z=0}$ and the synthetic electric field determined by the given membrane form.

### 3. Induced current pattern as graphene electromagnetic response for weak fields

If the external field $\vec{E}^{ext}(t)$ is weak (and consequently $\vec{E}^{syn}(\vec{r},t)$ is weak) and $P_{0x,0y} \ll 1$, and $G_{x,y}(Q_x, Q_y) \approx 1$, we can consider instead of the nonlinear self-consistent equations (16)-(18) their linear approximations

$$\frac{d}{d\tau}P_{0x} + \frac{\gamma}{\omega}P_{0x} = \frac{e}{\omega p_F}\left[E_x^{ext}(\tau) + E_x^{syn}(\vec{r},\tau)\right] \qquad (19)$$

$$\frac{d}{d\tau}P_{0y} + \frac{\gamma}{\omega}P_{0y} = \frac{e}{\omega p_F}\left[E_y^{ext}(\tau) + E_y^{syn}(\vec{r},\tau)\right] \qquad (20)$$

where $\tau = \omega t$. It is clear (see (5),(6)) that

$$E_{x,y}^{ext}(\tau) + E_{x,y}^{syn}(\vec{r},\tau) = E_{0x,0y}\sqrt{1 + \left(D^{(x,y)}(x,y,\omega)\right)^2}\cos[\tau - \delta_{x,y}(x,y,\omega)], \qquad (21)$$

$$\delta_{x,y}(x,y,\omega) = \arctan D^{(x,y)}(x,y,\omega), \quad D^{(x,y)}(x,y,\omega) = E^0(\omega)I_{x,y}(x,y) \qquad (22)$$

So we can rewrite (19),(20) in the form

$$\frac{d}{d\tau}P_{0x,0y} + \frac{\gamma}{\omega}P_{0x,0y} = \frac{e}{\omega p_F}E_{0x,0y}\sqrt{1 + \left(D^{(x,y)}(x,y,\omega)\right)^2}\cos[\tau - \delta_{x,y}(x,y,\omega)], \qquad (23)$$

Solving this equation we obtain

$$P_{0x,0y}(x,y,t) = \frac{e}{p_F}\frac{E_{0x,0y}}{\sqrt{\omega^2 + \gamma^2}}\sqrt{1 + \left(D^{(x,y)}(x,y,\omega)\right)^2}\cos[\tau - \delta_0(\omega) - \delta_{x,y}(x,y,\omega)],$$

$$\delta_0(\omega) = \arctan\frac{\gamma}{\omega} \qquad (24)$$

And consequently (see (10))

$$j_{x,y} = 2\frac{g_s e v_F}{(2\pi\hbar)^2}p_F^2 P_{0x,0y} =$$

$$= g_s\frac{e^2}{2\pi h}\frac{v_F p_F}{\pi h}\frac{E_{0x,0y}}{\sqrt{\omega^2 + \gamma^2}}\sqrt{1 + \left(D^{(x,y)}(x,y,\omega)\right)^2}\cos[\omega t - \delta_0(\omega) - \delta_{x,y}(x,y,\omega)] \qquad (25)$$

Neglecting corrugations influence i.e. for flatland model we get from (25) the formula obtained in [33].

$$j_{x,y} = g_s \frac{e^2}{2\pi h} \frac{v_F p_F}{\pi h} \frac{E_{0x,0y}}{\sqrt{\omega^2 + \gamma^2}} \cos[\omega t - \delta_0(\omega)] \qquad (26)$$

Note that our formula (25) calculated taking into account corrugation influence as well as (26) calculated without it look different compared to usually used Drude formula

$$\sigma(\omega) = \frac{\sigma(0)}{1 + (\omega\tau)^2}$$

However when $(\omega/\gamma) \ll 1$ (which is true in terahertz range, see (18)) these formulae weakly differ since then $\sqrt{1 + (\omega/\gamma)^2} \approx 1 + 2^{-1}(\omega/\gamma)^2$. But in terahertz range where the main mechanism of decay is radiation loss (not because of weak nonelasticscattering with phonons emission) our theory is in agreement with the experiment [38] where the intra conductivity i.e. its real part in terahertz range was measured to be a constant not depending on $\omega$. Also it was measured in [38] that inter part i.e. actually its imaginary part is close to zero in terahertz range. So we see that our quasiclassic approximation neglecting imaginary part of conductivity in terahertz range do not contradict to the experiment [38]

If we assume that there is no radiation loss (i.e. $\gamma = 0$) we get the formula for currents found in [18].

$$\vec{j}_x \approx \sigma_0 E_{0x}[\cos \omega t + D^{(x)}(x,y) \sin \omega t], \quad \vec{j}_y \approx \sigma_0 E_{0y}[\cos \omega t + D^{(y)}(x,y) \sin \omega t] \qquad (27)$$

$$\sigma_0 = \frac{n_s e^2 v_F}{p_F \omega}$$

The formulae (26) and (25),(27) for the current density in the linear approximation describe current paths which are straight lines in flatland model ( see (26)) and are curved lines depending on the point $(x, y)$ (see (25),(27)) when corrugations are taken into account. It means that the graphene membrane curvature leads to the curving of the current paths, the total current through the section conserving the constant value. So there is an important difference from the usual percolation description where the origin of the curved current lines is due to the random potential. Formulae (25) describe current patterns when radiation loss is taken into account. The formulae (25), (27) show that the induced current paths are curved which is the influence of corrugations always present. It leads to the fact that actually the current density depends on the point which was observed in a number of experiments (see [39]).

Note that the flatland model formula (26) for a valley current taken with the opposite sign gives us the valley current in another valley (see [19]) so that total current is equal to zero. However if we take into consideration the corrugations influence (formulae (25) or (27)) we have symmetry breaking and the total current is not equal to zero. The corresponding formula was obtained in [18] without taking into account

radiative loss. If radiation loss is taken into consideration (see (25)) we obtain for the total current

$$j_{x,y}^{tot} = 2\sigma_0 \, E_{0x,0y} D^{(x,y)}(x,y,\omega) \sin \omega t \,, \qquad \sigma_0 = g_s \frac{e^2}{2\pi h} \frac{v_F p_F}{\pi h \sqrt{\omega^2 + \gamma^2}} \tag{28}$$

In other words we have

$$\sigma^{tot} = 2\sigma_0 D^{(x,y)}(x,y,\omega)$$

So the presence of corrugations leads makes the conductivity a function of the point (x,y). So if we know the surface form $z = h(x,y)$ we can predict the form of current pattern. And visa versa if we know from the experiment the current pattern we can restore the surface form. This result could be used for creating new method of corrugations imaging when we have the experimentally measured current patterns.

## 4. Discussion and conclusion

In the present paper we obtained the generalized nonlinear self-consistent equation describing the corrugated graphene membrane electro-magnetic response in linear on the field approximation taking into account the radiative loss. We did not take into consideration the direct e-e interaction between particles but considered the electrodynamics interaction between currents using self consistent field method [32-33]. Obtaining this equation we assumed that the graphene membrane surface form changes under the action of the external field i.e. we have "breathing" corrugations. This assumption was justified by experimental works (see [39], fig.2, fig.3). Assuming the external field to be weak enough to use the linear approximation we simplified this equation which allowed finding exact solution. The current pattern found for the given surface form $z = h(x,y)$ proved to depend on the point $(x,y)$ which we predicted in [18] as the influence of the always existing ripples. Note that after the publishing of [18] we found experimental papers (see for instance [35], fig.3) which showed that this dependence really exists. However we could not compare our formula with experimental current patterns as we do not know the value of phenomenological constants we introduced. Note that according to our solution the induced currents which are the graphene electro-magnetic response the amplitude is less due to the radiation loss and the extra phase appears which depends on the point. It is important that morphology and current maps generally speaking do not coincide. Figuratively one may say that charge current in such graphene membrane is not homogeneous in any cross-section and reminds the "turbulent stream" of very thin layer of charged liquid intermixed by many dephased and unequal electro mixers. In the present paper we studied in addition the role of radiative decay of charge carriers "overheated" by terahertz irradiation which may be more effective than standard relaxation mechanism and make the picture of spatial alternating current paths more exact.

Note that we determined the gauge field $\vec{A}(x,y)$ on the basis of the formula [11] which was obtained using atomic scale quantum approach. It is especially important because it was experimentally found that there are in graphene nanoscale-wavelengh size ripples [40],[41].

For qualitative comparison between our formula and experimental current patterns the more in detail experimental investigation using methods developed in [35] should be made to study the values of the phenomenological constants we introduced. However one can determine these parameters experimentally using methods developed in [35].

In case new more thorough experimental investigation allows showing that our formula describes the phenomenon well enough we could try to solve the inverse problem i.e. the imaging of the surface form on the basis of the observed current patterns. It could be interesting for applications.

The problem of theoretical description of current patterns in different experiments is very important. The experimental study of spatially resolved photocurrents has already been done in monolayer graphene, [42], and in several graphene nanoribbon devices, [43], (the last is especially important for understanding of the functioning of ultrafast photodetectors and mode-locked lasers where graphene is used as an important component). Also it was considered by the description of coherent transport in graphene heterojunctions, [44], and in a graphene Quantum Dots fabricated by Atomic Force Microscope Nanolithograthy, [45], in a number of cases quantum mechanical approach being absolutely naturally used. So we see that this sphere of investigations has a large perspective. The mentioned experiments can be qualitatively explained on the basis of the results obtained in the present paper in the framework of description of local currents on the basis of corrugations influence.

Imaging experiments show that current profiles in low dimensional and mesoscopic systems may be very complicated and in the current density may arise streamlines. Note that for instance stationary theory of weak localization in graphene where the arising of streamlines and the interference of time reversal backscattered current trajectories were taken into account(see[46]) and the variant of this theory without taking into consideration these circumstances but taking into account graphene double valley spectrum (see [47]) come to a little different conclusions. The case of non-stationary phenomenon theory is obviously still more difficult and is awaiting the attentive investigator. And by constructing such non-stationary theories in graphene we think that the formulae we obtained in [17-18] and in present paper could be useful.